\documentclass[prb,showpacs,twocolumn]{revtex4}

\usepackage{graphicx}
\usepackage{amsmath}
\usepackage{dcolumn}
\newcommand{\la}{\mathord{\langle}}
\newcommand{\ra}{\mathord{\rangle}}

\newcommand{\beq}{\begin{equation}}
\newcommand{\eeq}{\end{equation}}
\newcommand{\bea}{\begin{eqnarray}}
\newcommand{\eea}{\end{eqnarray}}
\newcommand{\ds}{\displaystyle}
\newcommand{\bvec}[1]{\kern-.14em{\vec{\kern+.14em\bf#1}}}

\begin{document}

\title{Two-dimensional Heisenberg antiferromagnet in a transverse field}
\author{P.J. Jensen}
\author{K.H. Bennemann}
\affiliation{Institut f\"ur Theoretische Physik, Freie
  Universit\"at Berlin, Arnimallee 14,
  D-14195 Berlin-Dahlem, Germany}
\author{D. K. Morr}
\affiliation{Department of Physics, University of Illinois at
  Chicago, Chicago, IL 60607}
\author{H. Dreyss\'e}
\affiliation{IPCMS -- GEMME, Universit\'e Louis Pasteur, 23, rue du
  Loess, F-67037 Strasbourg, France }

\date{\today}

\begin{abstract}
We investigate the magnetic reorientation in a two-dimensional
anisotropic antiferromagnet due to a transverse magnetic field.
Using a many-body Green's function approach, we show that the
magnetization component perpendicular to the applied field (and
along the easy-axis of the antiferromagnet) initially increases with
increasing field strength. We show that this unexpected result
arises from the suppression of quantum and thermal fluctuations in
the antiferromagnet. Above the N\'eel temperature, this effect leads
to a reappearance of a magnetic moment along the easy-axis.
\end{abstract}


\maketitle

\section{Introduction} \label{sec:intro}
The magnetic properties of two-dimensional (2D) antiferromagnets
(AFM) have been extensively studied in the past.
\cite{TaI91,IKK99,CRT03,ZhN98,ZhC99,PMB99} Examples of 2D
AFMs that have been recently investigated include the manganites
which exhibit a colossal magnetoresistance,\cite{JTC94} the
vanadates,\cite{MAT01} and the undoped parent compounds of the
high-temperature superconductors.\cite{BGJ88} The latter are a
prime example of weakly anisotropic 2D AFMs due to their small
in-plane anisotropy and an even smaller interlayer coupling between
neighboring CuO$_2$-planes. A finite anisotropy is necessary to
stabilize the long-range magnetic order in 2D magnets at finite
temperatures.\cite{MeW66} Even a rather weak anisotropy induces an
ordering temperature of the same magnitude as the isotropic exchange.
\cite{Lin64} The properties of the above materials have been
intensely studied theoretically within the framework of the
Heisenberg model.
\cite{TaI91,IKK99,CRT03,BGJ88,DmK04,HSL05,BBG98,Din90}
In particular, the magnetization of the \textit{isotropic}
2D AFM as function of an applied magnetic field have been studied
within spin-wave theory.\cite{ZhN98,ZhC99,PMB99}

In this communication we study the properties of a 2D \textit{anisotropic}
AFM with spin $S=1/2$ on a square lattice in a \textit{transverse
magnetic field} perpendicular to the easy axis of the anisotropy. To
this end, we develop a many-body Green's function method
\cite{Tya67} that is based on the equation-of-motion formalism. Our
results are two-fold. First, the staggered AFM magnetization along
the easy axis \textit{increases} with increasing strength of the
transverse magnetic field. This effect is quite unexpected since the
field is directed perpendicular to the magnetization component. We
find, however, that the presence of a weak transverse field leads to
the suppression of quantum and thermal fluctuations in the AFM, the
former being responsible for the decrease of the zero temperature
sublattice magnetization from its saturation value, $S=1/2$. While a
similar behavior has been predicted for one-dimensional (1D) AFM
spin chains,\cite{DmK04} its relation to our results is at present
unclear due to the qualitatively different nature of spin
excitations in 1D and 2D systems. Second, the AFM ordered spins take a
noncollinear canted orientation for any non-zero transverse field,
inducing a non-zero magnetization component parallel to the applied
field. Note that this behavior is qualitatively different from that
of a 2D AFM in a longitudinal magnetic field parallel to the easy
axis.\cite{CRT03,HSL05} In the latter, a canted spin configuration
can only be reached via a phase transition into the so-called
spin-flop-phase.

\section{Theory} \label{sec:theory}
We consider the anisotropic (XXZ-) Heisenberg Hamiltonian
\begin{equation}
\label{e1} \mathcal{H} = \frac{1}{2}\sum_{\la ij\ra}\,\Big[J\;
\mathbf{S}_i \, \mathbf{S}_j + D\,S_i^z \, S_j^z \Big] -B_x \,
\sum_i \, S_i^x \,,
\end{equation}
where $\mathbf{S}_i$ is the spin operator with spin quantum number
$S=1/2$ located on sites $i$ of a square lattice, and $J>0$ is the
isotropic exchange coupling between nearest neighbor (nn) spin pairs
$i$ and $j$. The easy-axis is modeled by an exchange anisotropy
$D>0$ along the $z$-axis. We take the transverse magnetic field
$\mathbf{B}=(B_x,0,0)$ to be aligned along the $x$-axis,
perpendicular to the easy axis. Similar results to the ones
discussed below are expected for a single-ion anisotropy, although
its physical origin differs from the exchange anisotropy considered
in this study.\cite{FrK03}

In order to compute the temperature- and field-dependence of the
AFM's sublattice magnetization $\mathbf{m}_i(T,B_x)$, we employ a
many-body Green's function approach, which is based on the
equation-of-motion formalism.\cite{Tya67} The non-collinear
magnetic structure which occurs in the presence of a non-zero
transverse magnetic field requires that two non-vanishing
magnetization components have to be considered. This can be done
either in the original spin coordinate system,\cite{FJK00} which
however is analytically and numerically demanding due to the
occurring zero-eigenvalue problem.\cite{FrK05} Therefore, we apply
an approach identical to the one recently used to investigate the
field-induced spin reorientation of 2D anisotropic ferromagnets,
\cite{SKN05,PPS05} and to study the noncolllinear magnetization
of 2D isotropic antiferromagnets.\cite{ZhN98,ZhC99,PMB99}
In this approach, the spins of each sublattice
$i$ are rotated locally by angles $\theta_i$ in such a way that in
the rotated frame (primed spin operators) only a \textit{single}
non-vanishing component of the sublattice magnetization remains,
i.e., $m_i(T)=\la{S_i^z}'\ra\neq0$ and
$\la{S_i^x}'\ra=\la{S_i^y}'\ra=0$. The symmetry of the present case
simplifies the calculation considerably by assuming equal magnitudes
of the sublattice magnetization, $m(T,B_x)=|{\bf m}_1(T,B_x)|=
|{\bf m}_2(T,B_x)|$, and canting angles $\theta_1(T)=\theta(T)$ and
$\theta_2(T)=\pi-\theta(T)$ with respect to the easy axis.

Specifically, we consider the following commutator Green's functions
in energy space, \beq \label{e2} G_{ij}^\pm(\omega) =
\la\la{S_i^\pm}';{S_j^-}' \ra\ra_\omega \,, \eeq which we compute by
the conventional equation-of-motion approach. We approximate
higher-order Green's functions by using the Tyablikov decoupling for
$i \neq k$,\cite{Tya59} \beq \label{e3}
\la\la{S_i^z}'\,{S_k^\pm}';{S_j^-}'\ra\ra_\omega \sim
\la{S_i^z}'\ra\la\la{S_k^\pm}';{S_j^-}'\ra\ra_\omega =
m_i(T)\;G_{kj}^\pm(\omega) \;. \eeq For 2D ferromagnets with a small
anisotropy it has been shown that the Tyablikov decoupling (or
random-phase approximation, RPA) yields almost quantitative results
specifically for the magnetization and susceptibilities,\cite{HFK02}
whereas the resulting free energy and the specific heat are less well
described.\cite{JIR04} For systems with spin quantum number $S=1/2$
the magnetization can now be obtained from \beq \label{e4}
m(T)=1/2-\frac{1}{N}\;\sum_\mathbf{k}
\la{S_1^-}'{S_1^+}'\ra(\mathbf{k}) \;, \eeq where $N$ is the number
of lattice sites, and the momentum sum runs over the full Brillouin
zone. The equal-time correlation function
$\la{S_1^-}'{S_1^+}'\ra(\mathbf{k})$ is obtained after
Fourier transformation into momentum space from Eq.(\ref{e2}) via
the spectral theorem.\cite{Tya67} Note that here the indices $i,j=1,2$
refer to the two sublattices. We obtain
\bea  \label{e5}
\la{S_1^-}'\,{S_1^+}'\ra(\mathbf{k}) &=& \frac{m(T)}{2}\;\bigg[
\frac{a+b(\mathbf{k})}{\varepsilon_1(\mathbf{k})}\;
\coth\bigg(\frac{\beta\varepsilon_1(\mathbf{k})}{2}\bigg) \nonumber \\
&& \hspace*{-1cm}
+\frac{a-b(\mathbf{k})}{\varepsilon_2(\mathbf{k})}\;
\coth\bigg(\frac{\beta\varepsilon_2(\mathbf{k})}{2}\bigg) -2\bigg]
\;,  \eea where
\bea a &=& B_x\,\sin\theta(T) \nonumber \\
&+&\!\! q\,m(T)\,\big[J+D-(2\,J+D)\,\sin^2\theta(T) \big]
\,, \label{e6a} \\
b(\mathbf{k}) &=&
-m(T)\,\gamma(\mathbf{k})\,(J+D/2)\,\sin^2\theta(T) \,,
\label{e6b} \\
c(\mathbf{k}) &=&
m(T)\,\gamma(\mathbf{k})\,\big[J-(J+D/2)\,\sin^2\theta(T)
\big] \,, \label{e6c} \\
\varepsilon_{1,2}^2(\mathbf{k}) &=& [a\pm
b(\mathbf{k})]^2-c^2(\mathbf{k}) \,. \label{e6d} \\
\gamma(\mathbf{k}) &=& 2\,(\cos k_x+\cos k_y ) \,. \eea
Here, $q=4$ is the number of nearest neighbor sites,
and the lattice constant $a_0$ is set to unity. The excitations of
the system are represented by two branches of spin waves whose
dispersions are given by $\varepsilon_{1,2}(\mathbf{k})$.
For $\mathbf{B}=0, \theta=0$ the N\'eel temperature $T_N$ is given
by \bea  \label{e6e} T_N &=&
\frac{1}{4}\;\left(\frac{1}{N}\,\sum_\mathbf{k} \frac{\overline
a}{\overline a^2-(J\gamma(\mathbf{k}))^2}
\right)^{-1} \nonumber \\
&\approx& \pi\,J\Big/\ln\Big(\frac{J\,\pi^2}{q\,D}\Big) \;, \eea
with $\overline a=q\,(J+D)$. Note that the ``anomalous" Green's
function $G_{ij}^-(\omega)$ appears due to the canted nature of the
spin configuration in a transverse field. Its appearance implies
that in addition to the spin-flip term ${S^-}'{S^+}'$ also a term of
the form ${S^-}'{S^-}'$ is present in the
Hamiltonian.\cite{TaI91,Mil89} As a consequence, the spin precession
around the equilibrium direction is no longer spherical but
elliptical. Also, the consideration of $G_{ij}^-(\omega)$ guarantees
that the Mermin-Wagner theorem is fulfilled for an easy-plane magnet
($D<0$).

In order to obtain the individual components of the magnetization,
we need to compute the magnetization angle $\theta_0(T,B_x)$.
\cite{free} The latter can be obtained from the observation that in
the rotated frame in equilibrium no torque is exerted on
the magnetization $\la{S_i^z}'\ra$, i.e., ${S_i^z}'$ is a constant of
the motion, and $d\la{S_i^z}'\ra/dt=0$. This requirement, in
turn, is equivalent to the condition that the Green's function
$G_{ij}^z(\omega) = \la\la{S_i^z}';{S_j^-}'\ra\ra_\omega$ vanishes,
as has been shown for the field-induced spin reorientation of a
ferromagnetic monolayer.\cite{PPS05} Computing this Green's
function by using the same treatment as for the $G_{ij}^\pm(\omega)$-
Green's functions, i.e., by application of the Tyablikov decoupling,
one finds
\bea  \label{e7} && \hspace*{1cm} \sin\theta_0(T,B_x) =  \\
&&\hspace*{-0.5cm} \begin{cases} \frac{\ds B_x}{\ds
q\,m(T,B_x)\,(2J+D)}
\;,\quad & B_x<q\,m(T,B_x)\,(2J+D) \;, \\
\hspace*{1.5cm} 1 \;, & \hspace*{1cm} \mathrm{otherwise} \;.
\end{cases} \nonumber  \eea
This method to determine the angle $\theta_0(T,B_x)$ was
successfully applied by Schwieger {\it et al.}~\cite{SKN05} and by
Pini {\it et al.}~\cite{PPS05}\ for the case of the field-induced
spin reorientation of a 2D ferromagnet.

For $T=0$ we have compared this approach with the one where
$\theta_0(0,B_x)$ is determined from the minimum of the internal
energy $E(0,\theta)=\la\mathcal{H}\ra$, and found satisfactory
agreement. In fact, the relation Eq.(\ref{e7}) for $\theta_0(T,B_x)$
is identical to the one calculated from the free energy
$F_\mathrm{MFA}(T,\theta)$ within a single-site mean field
approximation. This approximation is easily obtained within our
theoretical approach by neglecting the spin-flip term $S_i^+S_j^-$
in Eq.(\ref{e1}), or putting $\gamma(\mathbf{k})=0$ in
Eqs.(\ref{e6b})-(\ref{e6d}).\cite{free}

\begin{figure}[t]
\includegraphics[width=5.2cm,angle=-90,bb=70 20 500 720,clip]
{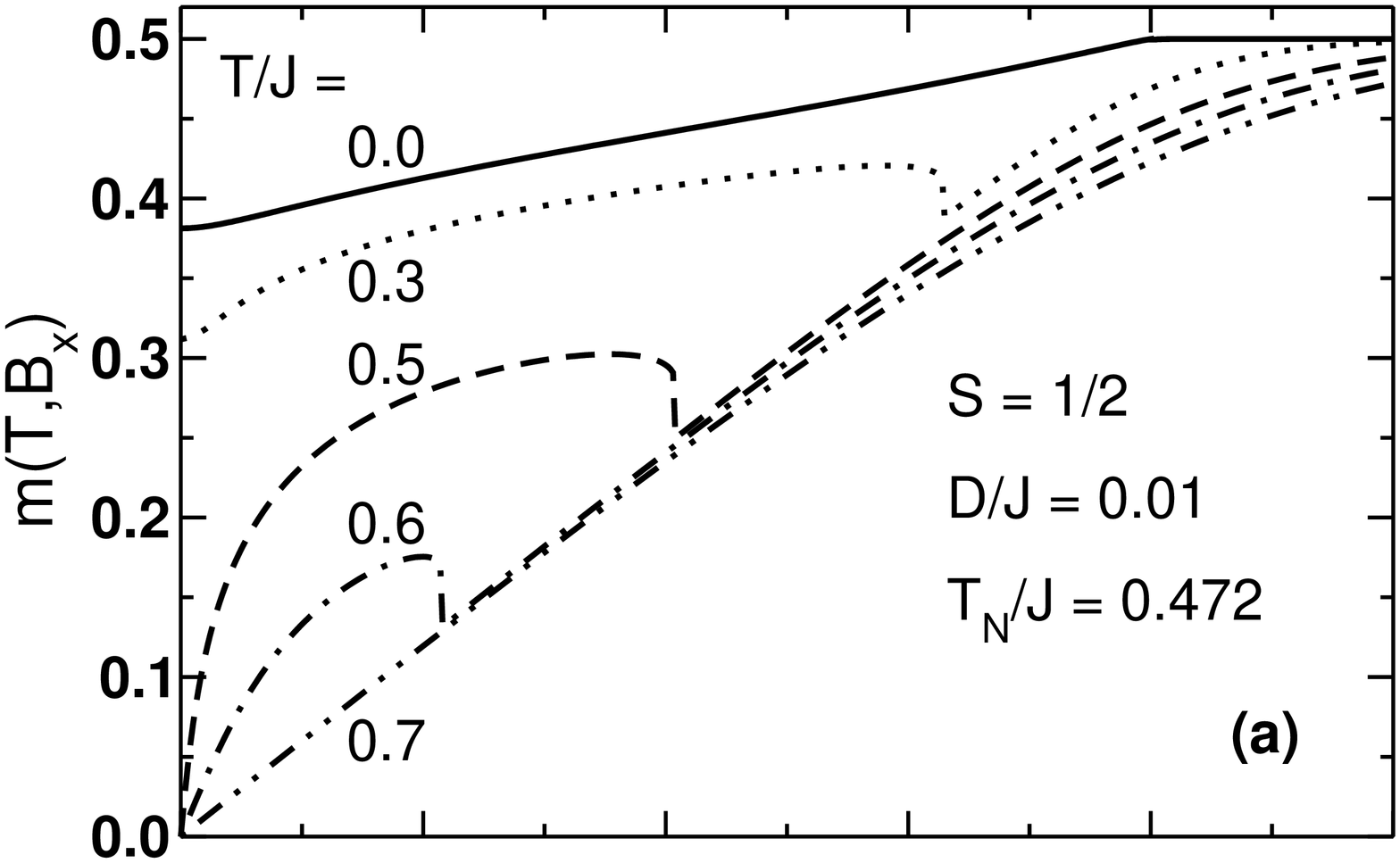} \\[0cm]
\includegraphics[width=5.8cm,angle=-90,bb=88 20 570 720,clip]
{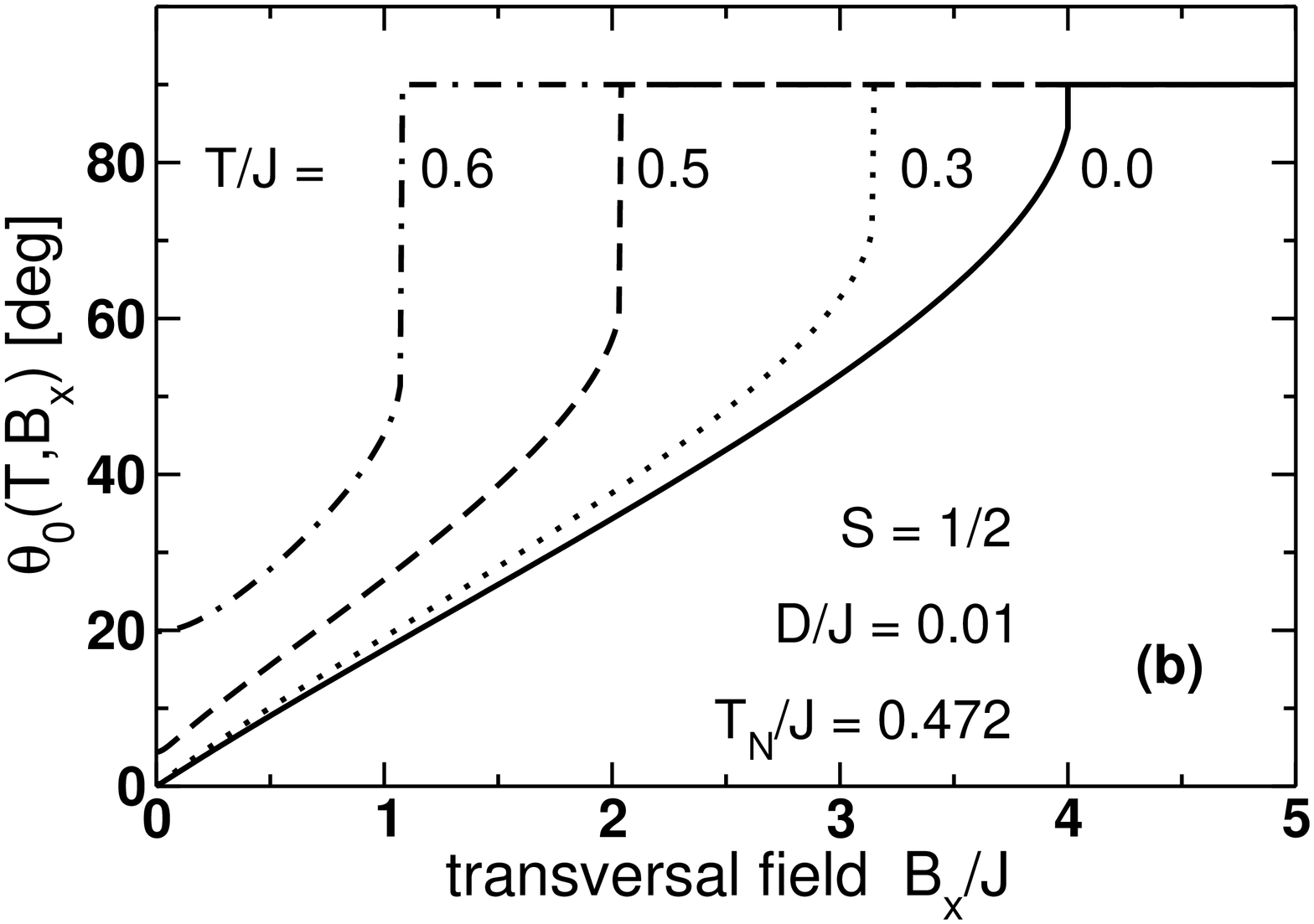}
\caption{(a) Magnetization $m(T,B_x)$ and (b) equilibrium
  angle $\theta_0(T,B_x)$ as functions of the transversal field $B_x$
  for different temperatures $T$ below and above the N\'eel
  temperature $T_N$. The temperatures and interactions are
  given in units of the isotropic exchange $J$. We have assumed
  a spin quantum number $S=1/2$ and an anisotropy $D/J=0.01$. }
\label{fig1} \end{figure}

\begin{figure}[h]
\includegraphics[width=5.8cm,angle=-90,bb=70 20 560 720,clip]
{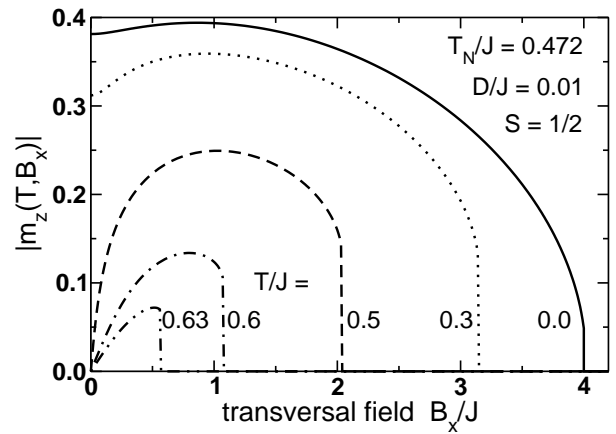}
\caption{Staggered magnetization component $|m_z(T,B_x)|$ along the
  easy axis as function of the transversal field $B_x$ for
  different temperatures $T$. For the other denotations we refer
  to Fig.~\ref{fig1}.} \label{fig2} \end{figure}

\section{Results} \label{sec:results}

Unless stated otherwise, we consider in the following a weak
anisotropy $D/J=0.01$, yielding a sublattice magnetization of
$m(T=0,B_x=0)=0.381$ at $T=0$. A magnetic anisotropy is
necessary since
for an isotropic system (i.e., for $D=B_x=0$), our approach
satisfies the Mermin-Wagner-theorem \cite{MeW66,Tya67} and we have
$m \equiv 0$ at $T \not = 0$. For the \textit{isotropic} 2D AFM
($D=0$) on a square lattice we obtain $m(T=0,B_x=0)\sim0.36$, which
compares reasonably well with the commonly accepted value of $0.307$.
\cite{BBG98} The latter value is also obtained within the
Holstein-Primakoff-approximation.
The N\'eel temperature $T_N$ for $D/J=0.01$ is
calculated to be $T_N/J=0.472$, while a quantum Monte Carlo
calculation yields a larger transition temperature of
$T_N/J=0.590$ for the same system.\cite{Din90}
In Fig.~\ref{fig1} we present the magnitude of the magnetization,
$m(T,B_x)$, and the equilibrium angle, $\theta_0(T,B_x)$, as a
function of the transverse field $B_x$ for temperatures $T$ below
and above $T_N$. For {\it all} temperatures, the
magnetization $m(T,B_x)$ increases when $B_x$ is increased from
zero. At the same time, $\theta_0(T,B_x)$ also increases with
increasing $B_x$, indicating that the magnetization is rotated
towards the direction of the transverse field. Moreover, for
$T<T_N$, the angle $\theta_0(T,B_x)$ deviates from zero for
infinitesimally small $B_x$, implying that the rotation of the
staggered moment does not require a critical field strength, in
contrast to the spin-flop transition associated with the application
of a longitudinal magnetic field.\cite{CRT03,HSL05} Note that for
$T>T_N$, the limit $B_x \to 0$ leads to a non-zero angle
$\theta_0(T,B_x)<\pi/2$. This implies that a non-zero $|m_z(T,B_x)|$
is induced by the transverse field, and that the component of the
induced magnetization parallel to the transverse field increases
faster than the component parallel to the easy axis.
For $T/J>0.64$ the behavior corresponds to the magnetization
of an isotropic AFM.\cite{ZhN98}

The reorientation field $B_R(T)=4m(T,B_R)(2J+D)$ is the smallest
field at which the magnetization is parallel to the direction of the
magnetic field, i.e., $\theta_0(T,B_R) = \pi/2$. Both $m(T,B_x)$
and $\theta_0(T,B_x)$ exhibit a discontinuous behavior at the
reorientation field. Specifically, $m(T,B_x)$ jumps at $B_R(T)$ to a
smaller value and increases with further increasing $B_x>B_R(T)$. At
the same time, the angle $\theta_0(T,B_x)$ jumps from $\theta <
\pi/2$ to $\pi/2$ at $B_R$.
Presently we cannot judge whether this discontinuous behavior is
`real' or an artefact of the approximations used above.

The staggered magnetization component $|m_z(T,B_x)|$ along the easy
($z$-) axis as computed from $m(T,B_x)$ and $\theta_0(T,B_x)$ is
shown in Fig.~\ref{fig2} as function of $B_x$. Though this
magnetization component is \textit{perpendicular} to the applied
field, we find that it also \textit{increases} when $B_x$ is
increased from zero. This behavior is quite unexpected since the
rotation of $\mathbf{m}(T,B_x)$ into the direction of the transverse
field should lead to a decrease in $|m_z(T,B_x)|$. After passing
through a maximum, $|m_z(T,B_x)|$ discontinuously vanishes at the
reorientation field $B_R(T)$. At the same time, the magnetization
component $m_x(T,B_x)$ along the magnetic field increases linearly
with $B_x$, and saturates for $B_x>B_R(T)$. As mentioned, for
$T>T_N$ a non-zero $B_x$ induces a finite $|m_z(T,B_x)|$ which
increases continuously from $|m_z|=0$ in the disordered
(paramagnetic) state for $B_x=0$, exhibits a maximum, and then
vanishes at the reorientation field. Moreover, in the limit $J=0$,
i.e., for a purely `Ising'-like exchange $D$ where the transversal
magnetization components are missing, we obtain that $|m_z(T,B_x)|$
does not exhibit a maximum, but decreases monotonically.

\begin{figure}[t]
\includegraphics[width=5.cm,angle=-90,bb=75 20 500 720,clip]
{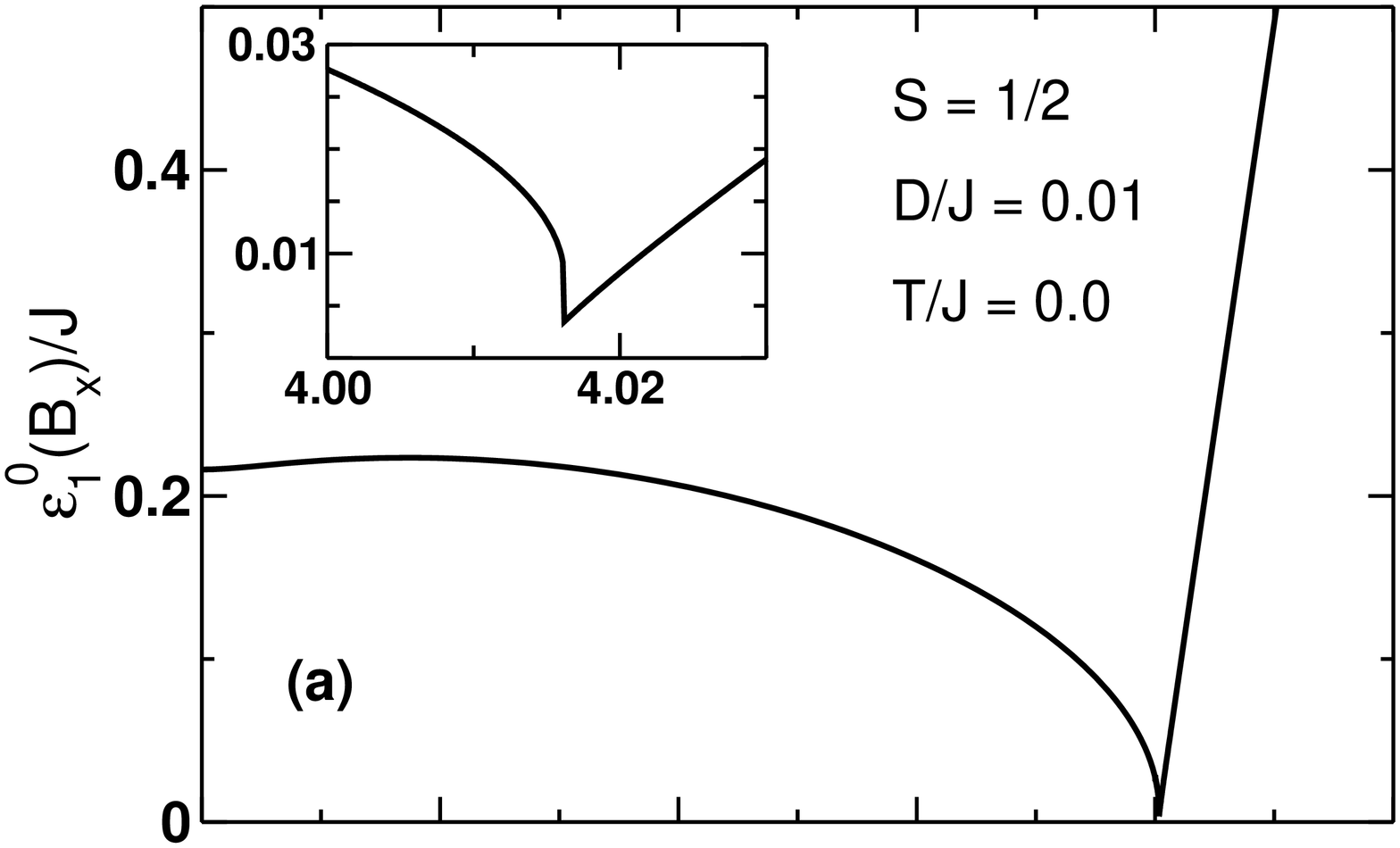} \\[0cm]
\includegraphics[width=5.65cm,angle=-90,bb=88 20 570 720,clip]
{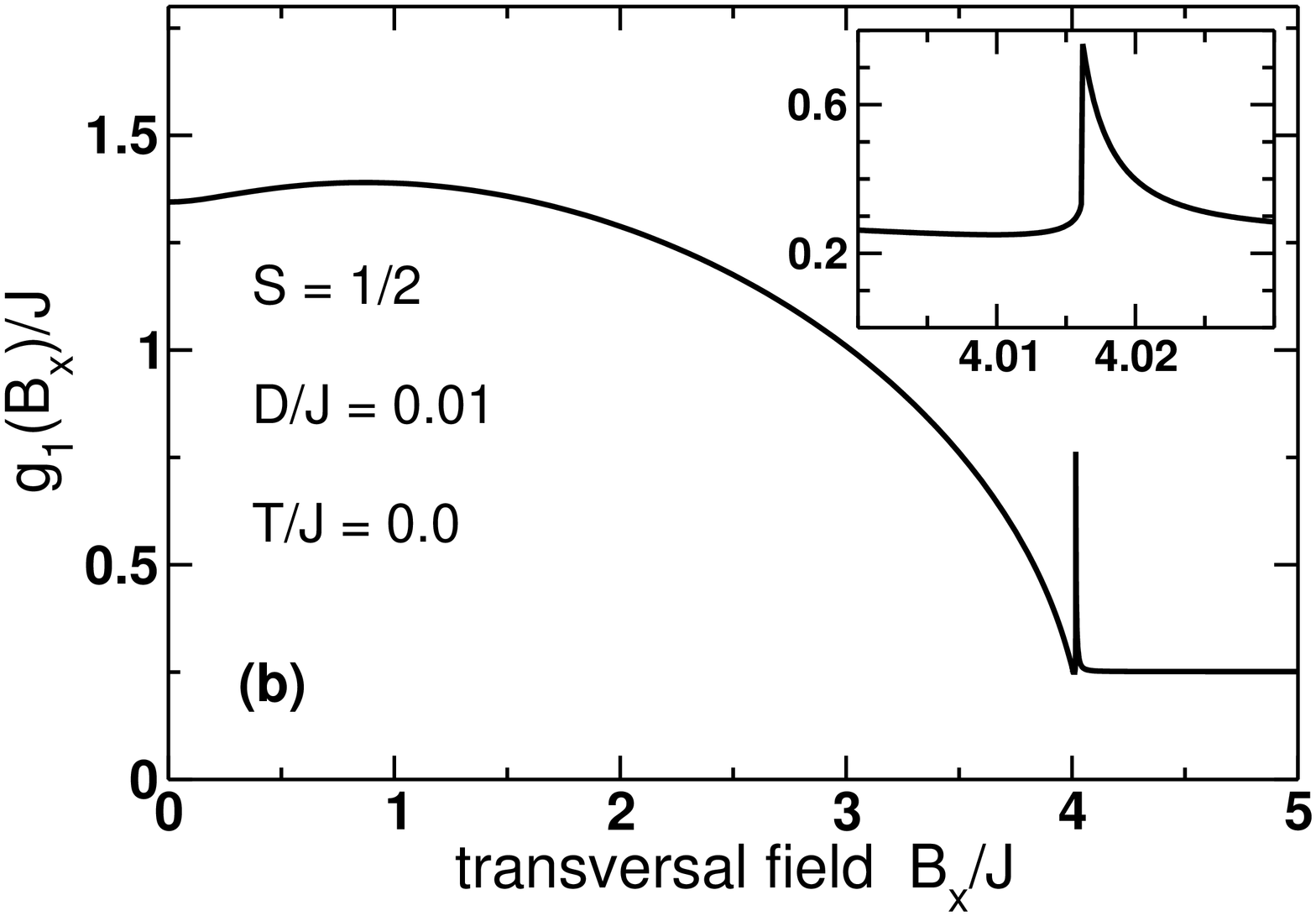}
\caption{(a) Energy gap $\varepsilon_1^0(B_x)$ and (b) spinwave
  stiffness $g_1(B_x)$, cf.\ Eqs.(\ref{e7},\ref{e9a}), of the magnon
  dispersion relation $\varepsilon_{1}(B_x)$ as function of the
  transversal field $B_x$ for $T=0$. The insets show the behaviors of
  $\varepsilon_1^0(B_x)$ and $g_1(B_x)$ near the reorientation
  field $B_R(0)/J \sim 4.0162$. For the other
  denotations we refer to Fig.~\ref{fig1}.}
  \label{fig3}\end{figure}

The increase of $|m_z|$ for small $B_x$ can be attributed to the
suppression of quantum and thermal fluctuations, whose strength
depends on the form of the magnon excitations spectrum. At $T=0$,
quantum fluctuations lead to a reduction of the staggered
magnetization $|m_z|$ from its saturation value given by $S=1/2$. In
order to investigate the strength of quantum fluctuations, we
consider the magnon dispersion, $\varepsilon_{1,2}(\mathbf{k})$ as a
function of the transverse field $B_x$ at $T=0$. Near the center of
the Brillouin zone at ${\bf k}=0$ we expand the dispersion to obtain
\beq \label{e8} \varepsilon_{1,2}(\mathbf{k})
=\varepsilon_{1,2}^0(B_x)+g_{1,2}(B_x)\,\mathbf{k}^2 \;, \eeq
where $\varepsilon_{1,2}^0$ is the energy gap (`mass') and $g_{1,2}$ the
spinwave stiffness of the dispersion. For fields smaller than the
reorientation field $B_R(0)$, we find
\bea \varepsilon_{1}^0(B_x) &=& \sqrt{ \frac{D}{2J+D}\;\Big[
(4m)^2\,(2J+D)^2 - B_x^2\Big]} \;, \label{e9a} \\
\varepsilon_{2}^0(B_x) &=& \sqrt{(4m)^2\,D\,(2J+D) + B_x^2} \;. \eea
In Fig.~\ref{fig3} we present $\varepsilon_{1}^0(B_x)$ and $g_{1}(B_x)$
for the lower-energy magnon branch.
Both quantities increase as $B_x$ is increased from
zero and exhibit a maximum at $B_x/J \sim 0.85$ which coincides with
the location of the maximum in $|m_z|$. The increase of the
excitation gap and of the spin stiffness reduce the strength of
fluctuations, and are thus directly responsible for the increase in
$|m_z|$. A decreasing anisotropy $D/J$ yields a smaller gap, which
increases the strength of the fluctuations, and as a result, the
maximum of $|m_z(0,B_x)|$ becomes more pronounced. For $B_x\to
B_R(0)$ the dispersion softens,\cite{PPS05} however, the gap
retains a small but still finite value at the reorientation field
$B_R(0)$. At the same time $g_1(B_x)$ exhibits a pronounced spike.
These features are responsible for the discontinuous behavior of the
magnetization at the reorientation field discussed above. Note that
for $T>T_N$ and for a finite $B_x$ the dispersions
$\varepsilon_{1,2}(T,B_x)$ do not become `soft' but exhibit as
expected a gap even when the magnetization is completely aligned
with the transverse field, i.e., for $\theta_0(T,B_x)=\pi/2$.

The increase of $|m_z(T,B_x)|$ at small $B_x$ is pronounced for
systems where quantum fluctuations are most important, i.e., for a
small spin quantum number $S$ and for a low spatial dimensionality.
Our calculations show that with increasing $S$ the relative maximum
of $|m_z(T,B_x)|$ becomes smaller, and is not present at all for
$S=\infty$, i.e., for classical spins.

\begin{figure}[h]
\includegraphics[width=6cm,angle=-90,bb=70 20 570 720,clip]
{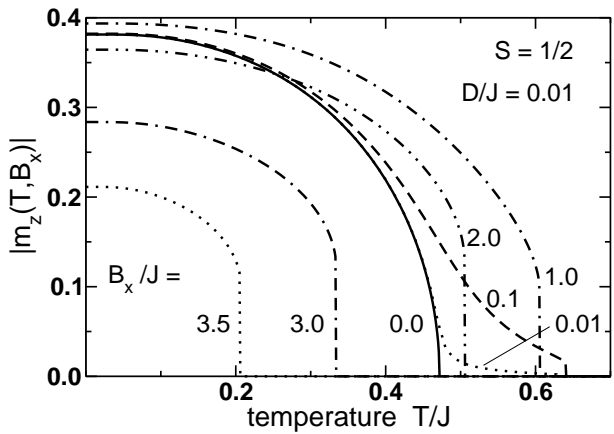}
\caption{Staggered magnetization component $|m_z(T,B_x)|$ along the
  easy axis as function of the temperature for different $B_x$. For
  the other denotations we refer to Fig.~\ref{fig1}. }
\label{fig4} \end{figure}

In Fig.~\ref{fig4}, we plot $|m_z(T,B_x)|$ as a function of the
temperature for various magnetic fields. We define the temperature
at which $|m_z|$ vanishes as the reorientation temperature
$T_R(B_x)$, with $T_R(B_x=0)=T_N$. Note that already
a weak magnetic field leads to a $T_R$ that is significantly
larger than $T_N$. With further increasing $B_x$ the reorientation
temperature decreases and vanishes for $B_x/J>4.02$.

In the remainder of this Section we will briefly mention related
results that were obtained in other systems and by using different
theoretical approaches. A maximum of $|m_z(T,B_x)|$ has recently
been obtained for an anisotropic AFM Heisenberg chain.\cite{DmK04}
Here $|m_z(T,B_x)|$ increases from the paramagnetic state $|m_z|=0$,
since such a chain does not exhibit an ordered state, and
corresponds thus to temperatures $T>T_N$ for the 2D AFM as
investigated in this study. Solving the latter system with a
single-spin mean-field approximation, no maximum of $|m_z(T,B_x)|$
is obtained, since the transversal spin terms
$S_i^x\,S_j^x+S_i^y\,S_j^y$ of Eq.(\ref{e1}), which cause the
obtained behavior, are neglected by this method. Nevertheless, if
these terms are taken into account to some extent, such as within a
two-spin mean-field approximation (Oguchi theory \cite{Ogu55}), the
properties of $|m_z(T,B_x)|$ are qualitatively reproduced. Finally,
a maximum of $|m_z(T,|B_z|,B_x)|$ is also obtained for an
antiferromagnetically coupled Heisenberg spin pair, a system which
can be solved exactly. In this case, for $B_x=0$ a finite magnetic
order is induced by a small staggered magnetic field $|B_z|$ along
the $z$-axis.

\section{Conclusion} \label{sec:conclusion}
We have studied the magnetization of a 2D square-lattice anisotropic
(XXZ-) AFM in a transverse magnetic field. A many-body Green's
function approach has been applied, which is known to yield a good
description of the magnetization for the case of a ferromagnetic
monolayer.\cite{JIR04} The fact that our theoretical approach
yields values for the sublattice magnetization at $\mathbf{B}=0$ and
the N\'eel temperature that are similar to those obtained in quantum
Monte Carlo calculations \cite{Din90} supports the validity of the
theoretical method also for the 2D AFM.

We showed that the staggered magnetization $|m_z(T,B_x)|$ along the
easy-axis perpendicular to the field increases for small $B_x$ and
exhibits a maximum before vanishing at the reorientation field. For
$T>T_N$, we demonstrate that the transverse field induces a non-zero
magnetization $|m_z(T,B_x)|$ which is perpendicular to the applied
field. We argue that the increase of $|m_z(T,B_x)|$ for small $B_x$
arises from changes in the magnon excitation spectrum, which in turn
leads to a suppression of thermal and quantum fluctuations.

The described behavior of the staggered
magnetization of a 2D AFM in a transverse field can possibly be
observed by, e.g., x-ray magnetic linear dichroism (XMLD), since
this method is sensitive to the magnitude of the magnetization
components.\cite{Laa98} An interesting question is whether a finite
magnetization component along the easy-axis for temperatures
slightly above $T_N$ as induced by a transverse magnetic field can
be measured, cf.\ Figs.~\ref{fig2},\ref{fig4}.

Note that typical magnetic fields of a few Teslas yield Zeeman
energies much smaller than the exchange, $B_x\ll J$. Hence, when
such a transverse field is applied to the AFM, the canting angle
will be small. In contrast, if the AFM is coupled to an ordered
ferromagnet (FM), the intrinsic field due to the strong interlayer
exchange coupling at the AFM/FM interface is considerably larger.
The resulting angle $\theta_0(T,B_x)$ could then be sufficiently
large such that the results presented above are observable. A
particular interest in such FM -- AFM interfaces has revived lately
in relation to the exchange bias effect.\cite{NoS99}

Useful discussions with K. D. Schotte are gratefully acknowledged.
P. J. J. likes to thank the IPCMS in Strasbourg, France, for the
hospitality, and the Deutsche Forschungsgemeinschaft, Sfb 290, for
financial support. D.K.M acknowledges financial support from the
Alexander von Humboldt Foundation.


\begin{thebibliography}{99}

\bibitem{TaI91} T. Tamaribuchi and M. Ishikawa,
    Phys. Rev. B \textbf{43}, R1283 (1991).

\bibitem{IKK99} % SSWT, review for comparable approx.
    V. Yu. Irkhin, A. A. Katanin, and M. I. Katsnelson,
    Phys. Rev. B \textbf{60}, 1082 (1999).

\bibitem{CRT03} A. Cuccoli, T. Roscilde, V. Tognetti, R. Vaia,
    and P. Verrucchi,  Phys. Rev. B \textbf{67}, 104414 (2003).

\bibitem{ZhN98} M. E. Zhitomirsky and T. Nikuni,
    Phys. Rev. B \textbf{57}, 5013 (1998).

\bibitem{ZhC99} M. E. Zhitomirsky and A.L. Chernyshev,
    Phys. Rev. Lett. \textbf{82}, 4536 (1999).

\bibitem{PMB99} D. Petitgrand, S. V. Maleyev, Ph. Bourges, and
    A.S. Ivanov, Phys. Rev. B \textbf{59}, 1079 (1999);
    A. V. Syromyatnikov and S. V. Maleyev,
    Phys. Rev. B \textbf{65}, 012401 (2002).

\bibitem{JTC94} % manganites
    S. Jin, T. H. Tiefel, M. McCormack, R. A. Fastnacht, R. Ramesh,
    and L. H. Chen, Science \textbf{264}, 413 (1994).

\bibitem{MAT01} % vanadates
    R. Melzi, S. Aldrovandi, F. Tedoldi, P. Carretta, % Li2VOSiO4
    P. Millet, and F. Mila, Phys. Rev. B \textbf{64}, 024409 (2001).

\bibitem{BGJ88} % exp and theor on doped LaCuO2
    R. J. Birgeneau et al., Phys. Rev. B \textbf{38}, 6614 (1988);
    E. Manousakis, Rev. Mod. Phys. \textbf{63}, 1 (1991).

\bibitem{MeW66} N. D. Mermin and H. Wagner, Phys. Rev. Lett.
    \textbf{17}, 1133 (1966).

\bibitem{Lin64} M. E. Lines, Phys. Rev. \textbf{133}, A841 (1964);
    M. Bander and D. L. Mills, Phys. Rev. B \textbf{38}, R12015 (1988);
    R. P. Erickson and D. L. Mills,
    Phys. Rev. B \textbf{43}, R11527 (1991).

\bibitem{DmK04} % Heisenberg chain, max of Mz in transversal field
    D. V. Dmitriev and V. Ya. Krivnov, Phys. Rev. B \textbf{70},
    144414 (2004).

\bibitem{HSL05} % classical 2D Heisenberg XXZ AFM in longitudinal field
    M. Holtschneider, W. Selke, and R. Leidl,
    Phys. Rev. B \textbf{72}, 064443 (2005).

\bibitem{BBG98} % QMC, M(0)
    B. B. Beard, R. J. Birgeneau, M. Greven, and
    U. J. Wiese,  Phys. Rev. Lett. \textbf{80}, 1742 (1998);
    J. K. Kim and M. Troyer,  Phys. Rev. Lett. \textbf{80}, 2705 (1998).

\bibitem{Din90} H. Q. Ding, J. Phys.: Condens. Matter \textbf{2}, 7979
    (1990); S. S. Aplesnin, Phys. Stat. Sol. B \textbf{207}, 491 (1998).

\bibitem{Tya67} S. V. Tyablikov, \textit{Methods in the quantum theory of
    magnetism\/,} Plenum Press, New York, 1967;
    W. Nolting, \textit{Quantentheorie des Magnetismus\/}, vol.2,
    B. G. Teubner, Stuttgart, 1986.

\bibitem{FrK03} P. Fr\"obrich and P. J. Kuntz,
    Europ. Phys. J. B \textbf{32}, 445 (2003).

\bibitem{FJK00} P. Fr\"obrich, P. J. Jensen, and P. J. Kuntz,
    Europ. Phys. J. B \textbf{13}, 477 (2000);
    P. Fr\"obrich, P. J. Jensen, P. J. Kuntz, and A. Ecker,
    Europ. Phys. J. B \textbf{18}, 579 (2000).

\bibitem{FrK05} P. Fr\"obrich and P. J. Kuntz,
    J. Phys.: Condens. Matter \textbf{17}, 1167 (2005).

\bibitem{SKN05} S. Schwieger, J. Kienert, and W. Nolting,
    Phys. Rev. B \textbf{71}, 024428 (2005).

\bibitem{PPS05} M. G. Pini, P. Politi, R. L. Stamps,
    Phys. Rev. B \textbf{72}, 014454 (2005).

\bibitem{Tya59} S. V. Tyablikov, Ukr. Mat. Zh. \textbf{11}, 287 (1959).

\bibitem{HFK02} P. Henelius, P. Fr\"obrich, P. J. Kuntz, C. Timm, and
    P. J. Jensen, Phys. Rev. B \textbf{66}, 094407 (2002).

\bibitem{JIR04} I. Junger, D. Ihle, J. Richter, and A. Kl\"umper,
    Phys. Rev. B  \textbf{70}, 104419 (2004).

\bibitem{Mil89} D. L. Mills, Phys. Rev. B \textbf{40}, 11153 (1989).

\bibitem{free} Usually the equilibrium angle $\theta_0(T,B_x)$ should
    be determined by minimizing the free energy
    $F(T,\theta)$ as obtained by the Green's function approach.
    Unfortunately, since within the Tyablikov decoupling $F(T,\theta)$
    behaves unphysically at elevated temperatures,\cite{JIR04} this
    approach cannot be applied here.

\bibitem{Ogu55} T. Oguchi, Progr. Theor. Phys. \textbf{13}, 148 (1955).

\bibitem{Laa98} G. van der Laan, Phys. Rev. B \textbf{57}, 5250 (1998);
    J. L\"uning, F. Nolting, A. Scholl, H. Ohldag, J. W. Seo,
    J. Fompeyrine, J.-P. Locquet, and J. St\"ohr, Phys. Rev. B
    \textbf{67}, 214433 (2003).

\bibitem{NoS99} For recent reviews see, e.g.: J. Nogu\'es and I. K.
    Schuller, J. Magn. Magn. Mater. \textbf{192}, 203 (1999);
    M. Kiwi, \textit{ibid.} \textbf{234/3}, 584 (2001);
    R. L. Stamps, J. Phys. D: Appl. Phys. \textbf{33}, R247 (2000).

\end{thebibliography}
\end{document}